\documentclass{article}
\usepackage[english]{babel}
\usepackage[T2A]{fontenc}
\usepackage[cp1251]{inputenc}
\usepackage[dvips]{graphicx}
\usepackage{amsmath}
\usepackage{amssymb}
\usepackage{latexsym}
\begin{document}
\rm
\let\sc=\sf
\begin{flushright}
Journal-Ref: Astronomy Reports, 2014, No. 1
\end{flushright}
\begin{center}
\LARGE {\bf Spectroscopy of $\varepsilon$ Aurigae during the 2009-2011 eclipse.}\\

\vspace{1cm}
\Large {\bf I. S.\,Potravnov$^{1,2*}$, V. P.\,Grinin$^{1,2}$}

\normalsize \vspace{5mm} 1 -  Pulkovo Astronomical Observatory,
Russian Academy of Sciences, 196140, Pulkovo, St.\,Petersburg,
Russia\\

2 - The Sobolev Astronomical Institute, St. Petersburg University,
Petrodvorets, St.
Petersburg, Russia\\
\end{center}

\normalsize
\begin{abstract}
The results of the spectral observations of a unique eclipsing
system $\varepsilon$ Aurigae obtained during the recent eclipse
are presented. We traced the spectral lines variations at
different eclipse phases. The part of these changes is the result
of the absorption of stellar radiation by the gas component of
circumstellar (CS) disk around the secondary. The other part
arises due to the obscuration of the stellar photosphere and the
closest circumstellar area by the dusty CS disk. Comparison of our
results with the previous one, obtained by Lambert \& Sawyer
during the 1982-1984 eclipse, indicates the good agreement, and
argues in favor of stability of the secondary disk structure. On
the base of the new observational data we provide a new mass
estimation of the system components.
\end{abstract}

  \vfill
\noindent\rule{8cm}{1pt}\\
{$^*$ e-mail: $<$ilya.astro@gmail.com$>$}

\clearpage

\section{Introduction}

$\varepsilon$ Aur is one of the most enigmatic eclipsing binary
system. Variability of this star was discovered by the pastor Fritch
in 1821. However, only in 1903 Ludendorf showed that it is the
eclipsing binary with the unusually long period: $\sim 27.1$
years. The duration of the eclipses is even more surprising:
approximately two years. Together with the long orbital period it
indicates that the size of the obscuring body is much larger than
the largest known stellar diameters. Also it means the precise edge-on
orientation of the orbital plane to the line of sight. The main
component of the system is the bright F0 Ia supergiant ($V\sim
3.0^m$). Every 27 years it is occulted by the secondary component
which does not manifest itself in the optical wavelengths range.
Currently, due to the substantial distance uncertainty only the
mass function of the system $\emph{f}\sim 2.5 M_\odot$ [1,2] is
known. Therefore the term 'main' traditionally applies to the
visible F star.

 Two different types of the models of $\varepsilon$ Aur were considered (e.g.
[3] and the papers cited there). In one of them, so-called,
'massive' model, the optical component is the normal supergiant
with the mass $M \sim 15M_ \odot$, which has left recently the
Main Sequence. In this case the secondary has the mass $M \sim
13M_ \odot$. In the second, so-called, 'low-mass' model, the main
component is an evolved post-AGB star with the mass close to 1
$M_\odot$; the secondary is more massive: $M \sim 5M_ \odot$. The
common assumption for both these models is that the
obscuring body is the CS disk of the secondary [4],
that was confirmed by the interferometric observations during the
last 2009-2011 eclipse [5].

The photospheric spectrum of the main component remains visible
during the all eclipse phases. The additional spectral lines arise
in the disk around the secondary and blend the spectrum of the
primary. These lines have been observed beginning from the early
spectroscopic investigations of the star. These lines appear in
the ingress phase and disappear after the end of the eclipse [6].
The spectra of $\varepsilon$ Aur obtained by Struve et al. [7]
during the 1954-1957 eclipse confirmed this and demonstrated the
complex variability of the spectral lines. Lambert \& Sawyer [8]
estimated the masses of the system components: $M_1\leq 3M_ \odot$
и $3 \leq M_2 / M_ \odot \leq 6$, using the K I 7699 \AA\ radial
velocity and the assumption that the secondary disk is the
Keplerian one. It was the argument in favor of the 'low-mass'
scenario.

During the 1982-1984 eclipse there were several attempts to
distinguish the spectral lines which are formed in the disk of the
secondary [9, 10]. It was shown that such lines had the
excitation temperature $T \sim 4000 K$ [10]. Thereby, in the spectrum of the primary
the lines with the high excitation potential such as
NI triplet, or the high members of the Paschen series do not
exhibit the eclipse effects, and their radial
velocities trace the orbital motion of the optical component. \\
Any attempts to observe the spectral lines of the secondary
component were unsuccessful [11].

\section{Observations and data reduction}
The last eclipse of $\varepsilon$ Aur occurred in 2009-2011. The
result of the long term observations of the radial velocity were
the updated orbital parameters and ephemeris [1, 2, 12].
Unfortunately, the argument of the pericenter is determined with
some uncertainty because it is impossible to detect the secondary
eclipse. The additional source of the uncertainty in the stellar
parameters estimate are pulsations of the primary (see below).
According to the results of the international observational
campaign [13], the mid-eclipse in the \emph{V} band occurred on
22.07.2010 (JD 2455400). The photometric moments of the eclipse
phases are presented in Table 1.

\begin{table}[h]
\begin{center}
\begin{tabular}{|c|c|c|}
\hline
Contact & Date & JD  \\
\hline
I & 16.08.2009 &2455070\\
II &22.02.2010 &2455250\\
III &27.02.2011 &2455620\\
IV  &26.08.2011 &2455800\\
\hline
\end{tabular}

\end{center}
\caption{\label{Table 1} The moments of the photometric contacts
[13]}
\end{table}

The spectra discussed in our paper were obtained using the high
resolution (R = 45000) echelle spectrograph MAESTRO with the 2 m
telescope at the Terskol Observatory. We obtained 58 spectra in
the 3900-9800 \AA\ wavelength range during the time range from
03.04.2009 to 30.11.2012. Our observations cover practically all
eclipse phases except the moment of egress, when observations
could not be fulfilled due to the technical problems. The
comparison between the out of eclipse spectra and the spectra
obtained during the eclipse provide an opportunity to trace the
spectral changes arising due to the passage of the secondary's
circumstellar disc across the line of sight. The standard
procedure of the spectroscopic data reduction was fulfilled with
the IRAF software package [14]. The wavelength calibration has
been done using the Th-Ar lamp comparison spectra obtained during
each observation night. The heliocentric correction was applied.
The continuum normalization of spectral orders has been done using
the spline approximation of the manually determined points. This
allows us to take accurately into account the broad wings of some
spectral lines. The further analysis of the spectra, including the
measurement of the radial velocities and lines equivalent widths,
has been done using the DECH 30 software [15].

\section{Results}
\subsection{ KI 7699 \AA\ line}
\begin{figure}[h] \centering
\includegraphics[width=6cm,angle=0]{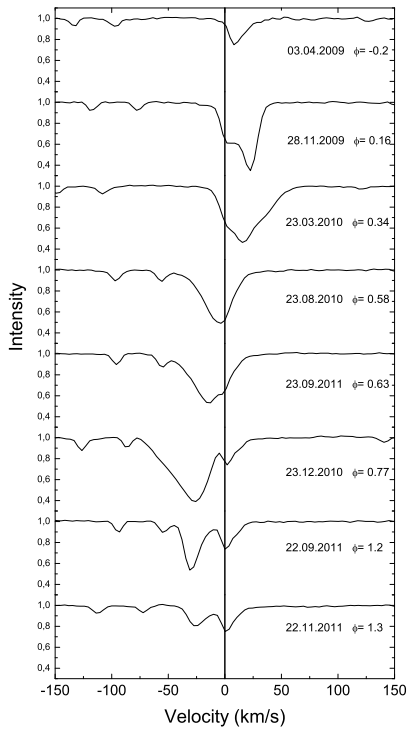}
\caption{The KI 7699 \AA\ line profiles variability during the
eclipse. Two weak absorption lines at the blue part of plots are
the telluric lines. Their shift is the result of the heliocentric correction.} \label{1}
\end{figure}

The resonance potassium doublet KI 7664 \AA\ and 7699 \AA\ absent
in the photospheric spectrum of the main component. The weak lines
of this doublet which we observed in the out of eclipse spectra
had the interstellar origin [16]. Nevertheless, these lines
demonstrate the strong variability during the eclipse. This
variability arises due to scattering of the photospheric radiation
by the neutral potassium atoms in the disk of the secondary.
Therefore, the investigation of this variability reveals the
dynamics of the gas component in the disk. A comprehensive
analysis of the KI 7699 \AA\ variability during the 1982-1984
eclipse presented by Lambert \& Sawyer [8]. Unlike
to the stronger KI 7664 \AA\ line, this line is less affected by the
telluric $O_{2}$ spectrum. Therefore, for our analysis we
chose the KI 7699 \AA\ line. Figure \ref{1}
demonstrates the consequence of changes in this line during the
eclipse.  For simplicity, we show
only part of the spectral line profiles for selected eclipse
phases.

The X-axis is the phase of the eclipse $\phi$. The zero point is
the moment of the first photometric contact (duration of the
eclipse is accepted as a unit). On the first out of eclipse
spectrum, the weak interstellar potassium line is clearly seen.
However, just after the beginning of the eclipse the additional
strong red-shifted absorption has appeared. On the following
spectra this absorption became stronger and gradually merged with
the interstellar component. Close to the mid-eclipse phase the
profile of this blend demonstrates a small asymmetry of the blue
wing, which evolves into the separate absorption component. In the
last two profiles obtained at the end of the eclipse this
additional absorption became weaker.

\begin{figure}[h] \centering
\includegraphics[width=8cm,angle=0]{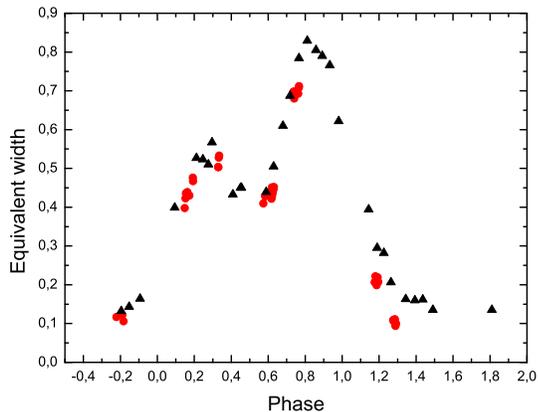}
\caption{Variability of the KI 7699 \AA\ line equivalent width
during the eclipse (red circles). The data by Lambert \& Sawyer
[8] are marked by triangles.} \label{2}
\end{figure}

Variability of the KI 7699 \AA\ line equivalent width (EW(KI))
is shown in Fig. \ref{2}. The dependence of EW(KI) on
the phase of eclipse reaches the maximum near the second and
third contacts. There is a difference between an amplitude of
these maxima: it is smaller near the second contact. Such an
asymmetry appears also on the radial velocity curve (see below).
The variability of EW(KI) arises because of the
variations of the potassium atom column density in the CS disk of
the secondary in the projection on the primary's disk. The
comparison of our data with the data obtained by Lambert \& Sawyer
[8] during the previous eclipse shows the good agreement.
According to their assumption, the difference in the EW(KI)
amplitudes at the ingress and egress phases is due to the
non-symmetric position of the central body in the circumstellar
disk.

\begin{figure}[h] \centering
\includegraphics[width=8cm,angle=0]{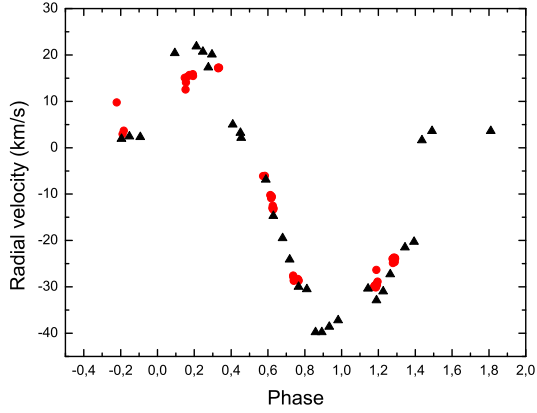}
\caption{Behavior of the KI 7699 \AA\ radial velocities during the
eclipse (red circles). The data by Lambert \& Sawyer [8] are
marked by triangles.} \label{3}
\end{figure}

Figure \ref{3} demonstrates the KI radial velocity curve (RV(KI))
during the eclipse. We also see here the good agrement between the
present and the previous RV curves. This argues in favor of the
stability of the secondary circumstellar disk. When intersecting
the line of sight, the rotating disk absorbs the radiation of the
primary firstly at the red side of the interstellar line, and then
at the blue side. At the second part of the eclipse an interesting
detail is clearly seen (Fig. 1): an additional blue shifted
absorbtion component appeared. It is save till the last phase of
observations (November 2012 ) and possibly later. This indicates
an existence of the gaseous matter on the line of sight after the
end of the photometric eclipse. This matter may be the extended
gaseous structure like a stream trailing behind the disk of the
secondary. Such structures can appear in the evolved binary
systems (see e.g. [17]).

\subsection{The D Na I resonance lines}

\begin{figure}[h] \centering
\includegraphics[width=7cm,angle=0]{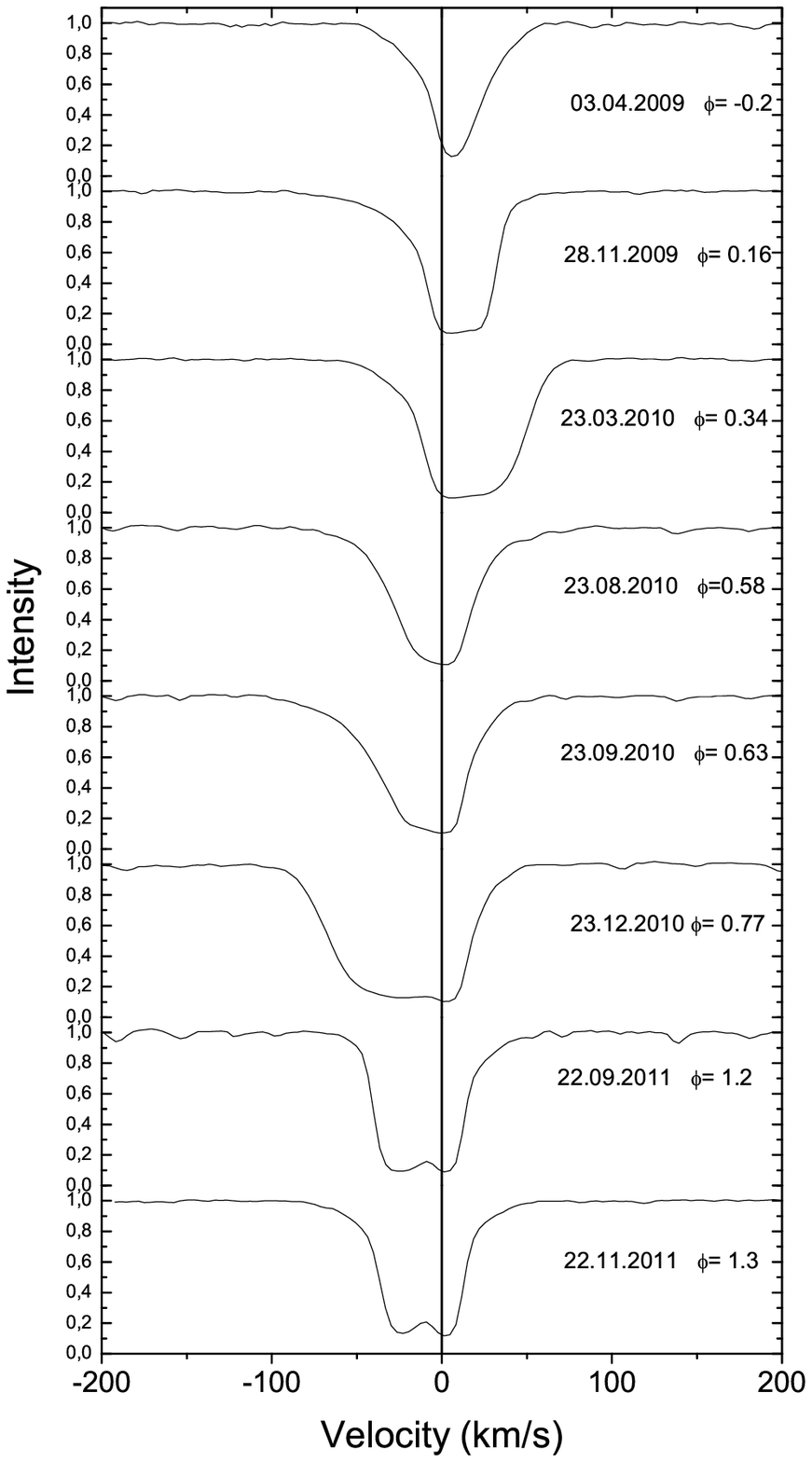}
\caption{Changes in the NaI 5895 \AA\ line profiles during the
eclipse.} \label{4}
\end{figure}

The sodium doublet 5889 \AA\ and 5895 \AA\ in $\varepsilon$
Aurigae spectrum demonstrate a complex structure: the profiles
consist of both interstellar and weak stellar components which
originate in the photosphere of the primary. The contribution of
the first component is more significant. Therefore, the sodium lines
do not demonstrate any noticeable variability due to the primary
pulsations. The spectral variability
 in the Na I lines during the eclipse (Fig. \ref{4}) was similar to that observed in the KI 7699 \AA\ line. Due to the low
excitation potentials, additional absorption components observed
during the eclipse were formed at the similar physical conditions. In
particular, as in the case of the KI 7699 \AA\ line, the
additional blue-shifted absorption component of the D NaI lines
appeared after the middle of the eclipse and was observed after
the end of the eclipse.

\subsection{The FeII 4515 \AA\ line}

\begin{figure}[h] \centering
\includegraphics[width=8cm,angle=0]{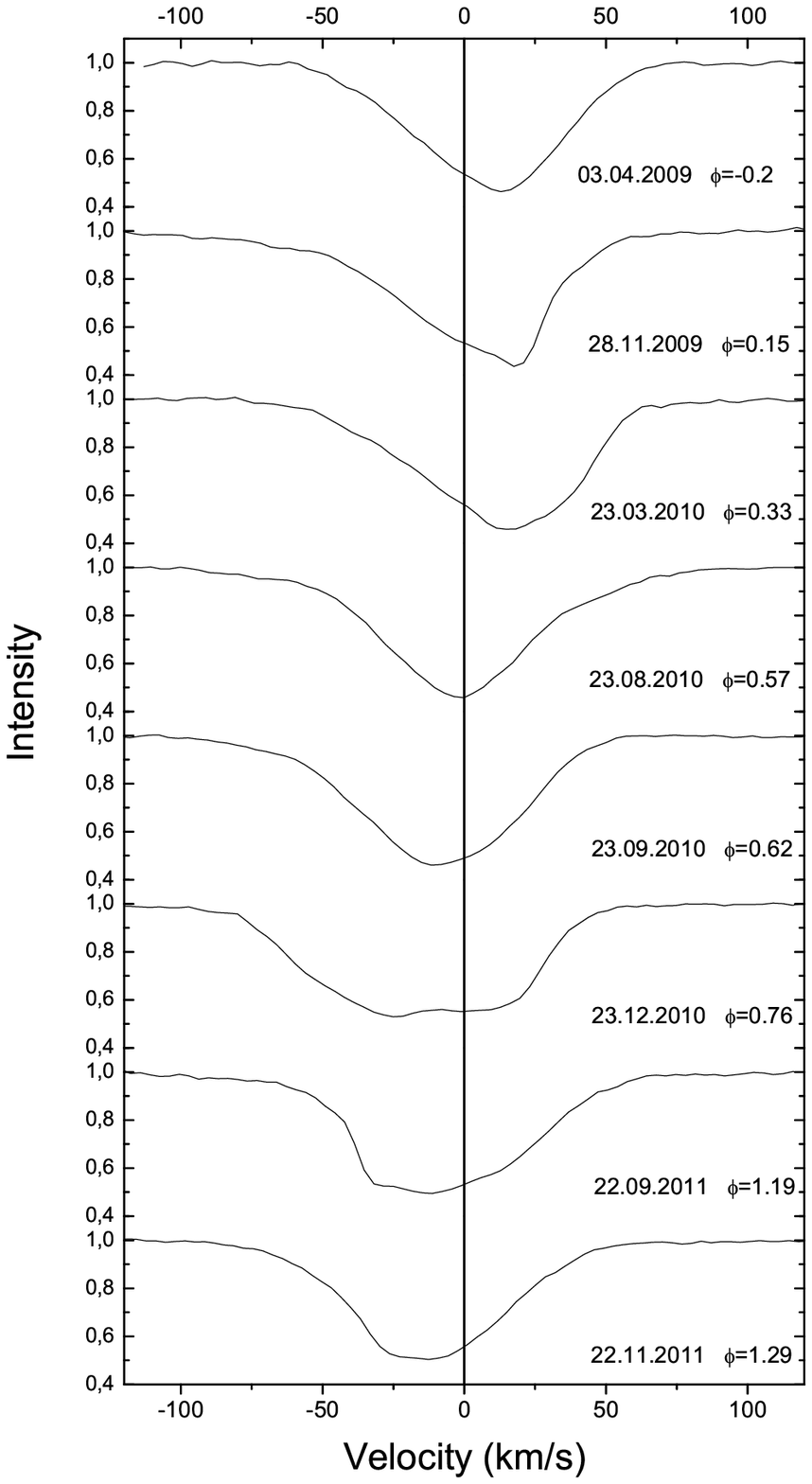}
\caption{FeII 4515 \AA\  line profiles variability during the
eclipse.} \label{5}
\end{figure}

\begin{figure}[h] \centering
\includegraphics[width=8cm,angle=0]{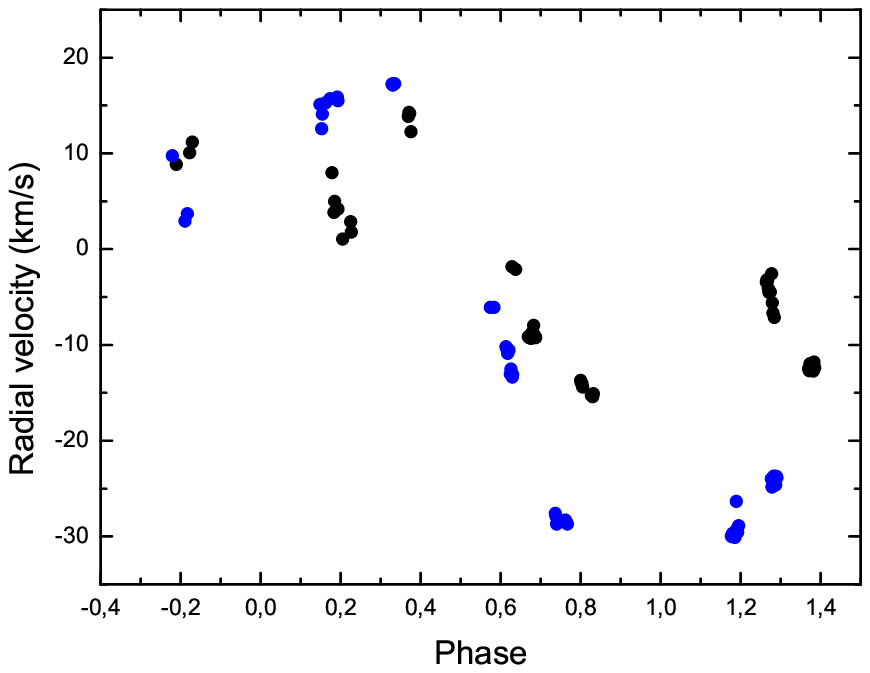}
\caption{The radial velocities of the FeII 4515 \AA\ (black
circles) and KI 7699 \AA\ (blue circles)  lines during the
eclipse.} \label{6}
\end{figure}

The lines of the ionized iron are widely presented in the spectrum
of $\varepsilon$ Aur. For an analysis we chose one of them,
namely, the unblended FeII 4515 \AA\ line. In general, the
variability of this line is in a good agreement with the
previously described changes in the alkali metals line but in a
smaller scale. The appearance of the red- and blue-shifted
additional absorption is clearly traced in the ingress and egress
phases correspondingly. Figure \ref{5} presents the consequent
changes in the line profile during the eclipse. In Fig. \ref{6} we
compare the radial velocity curves both in the FeII 4515 \AA\ and
KI 7699 \AA lines. One can see a good general agreement in the
behavior of these curves. However, some lag is clearly seen. At
the phase $\phi\sim 0.2$ the KI radial velocity begins to grow due
to the appearance of the additional red-shifted absorption. At the
same time, the FeII radial velocity is still close to the orbital
motion velocity of the F star. The additional absorption in this
line appears at the later phases of the eclipse, and disappears
earlier. Near the end of the eclipse, the RV amplitude in the FeII
4515 \AA\ line is lower than in the KI 7699 \AA. After the end of
the eclipse the radial velocity in the iron line quickly returns
to the orbital one. This means that the formation region of this
line is more compact in comparison with the KI line formation
area.


\subsection{The CaII K line}
\begin{figure}[h] \centering
\includegraphics[width=8cm,angle=0]{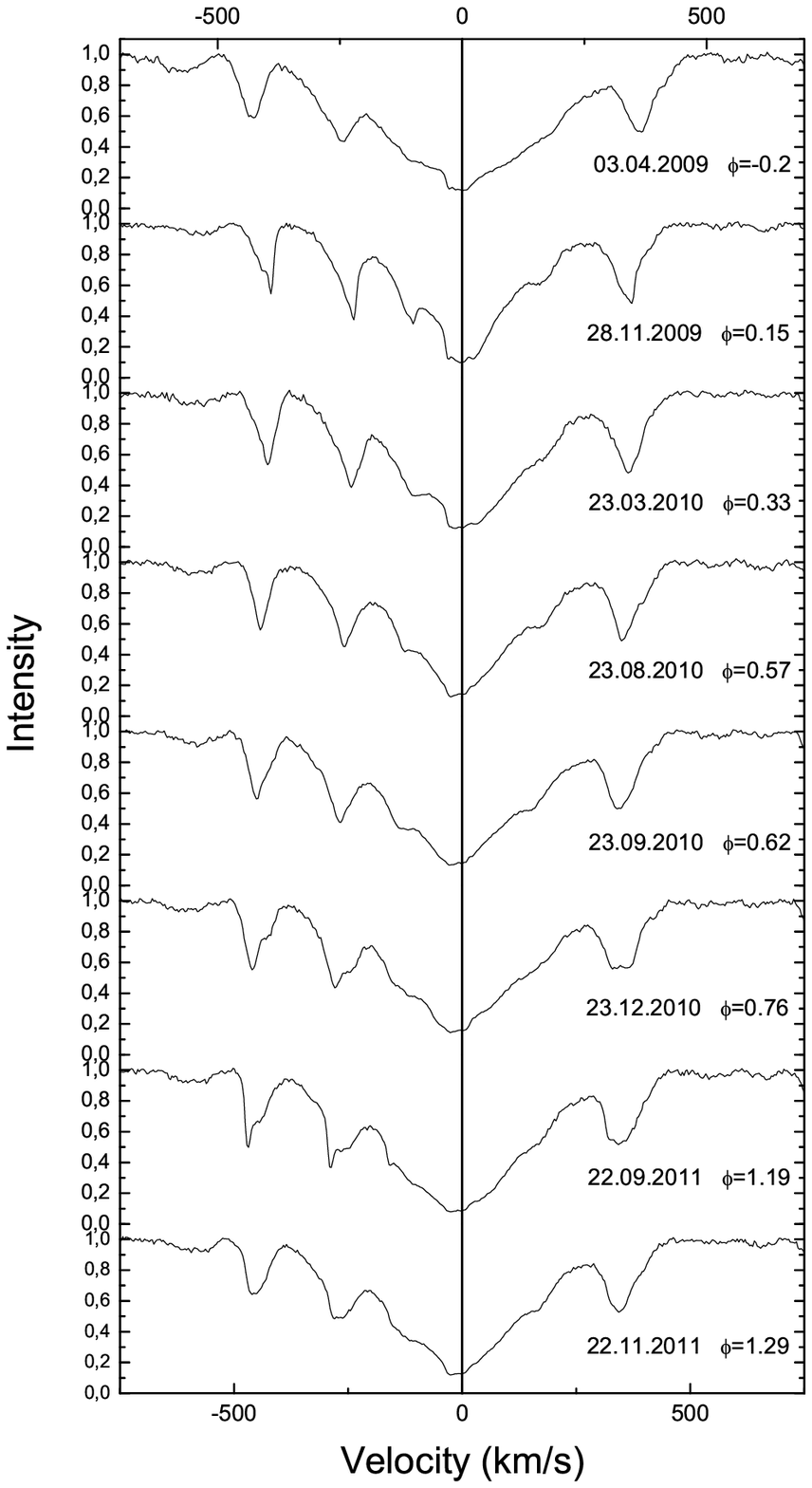}
\caption{CaII K line profiles at the different eclipse phases.}
\label{7}
\end{figure}

The CaII K resonance line proved to be quiet stable line in the
$\varepsilon$ Aur spectrum. The line profile looks practically the
same at all phases of the eclipse (Fig. \ref{7}), and do not
demonstrate any noticeable additional absorption associated with
the disk of the secondary. This means that there is no enough amount of
the calcium atoms in the disk around the secondary, which could
cause the changes in the line profile during the eclipse. This
fact is really surprising because this line is the resonance one, and
calcium is one of the most abundant metal.

\subsection{The NI and P 12 lines}

\begin{figure}[H] \centering
\includegraphics[width=8cm,angle=0]{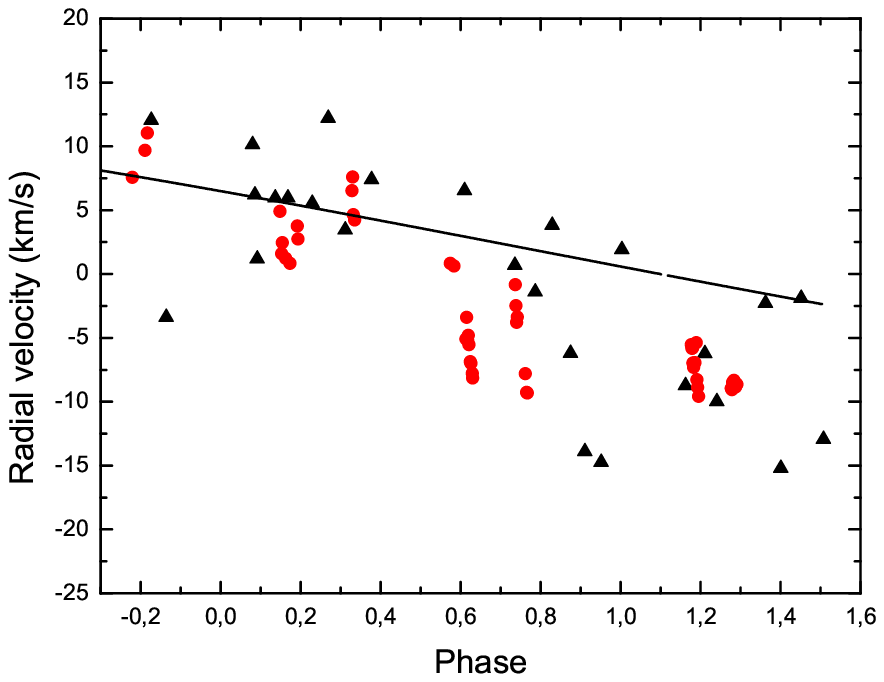}
\caption{The variability in the NI 8711\AA\ line radial velocity
(red circles). The data by Lambert \& Sawyer [8] are shown by
triangles. The orbital motion of the primary was calculated on the
base of the orbital solution [2] and shown by the line.} \label{8}
\end{figure}

Due to the high excitation potentials, we allocated the near-infrared nitrogen
triplet NI 8703, 8711, 8718 \AA\ and hydrogen Paschen 12 line in the separate group. These lines are formed in
the photosphere of primary and do not demonstrate any effects caused
by the eclipse. Thus, as mentioned above, the radial
velocities measured in these lines, traced the orbital motion of
the F star. Nevertheless, the nitrogen lines demonstrate a small
variability. The gradual intensification observed in blue and
red wings of the NI lines was not correlated with the eclipse phases.
The similar changes in these lines has been observed by Lambert \&
Sawyer [8] during the previous eclipse. Such a variability is
probably caused by the pulsations of $\varepsilon$ Aur. It should
be noted that this effect directly influences the precision of
the radial velocity measurements. For the direct measurements of
RV we choose the nitrogen line 8711 \AA. Figure \ref{8} shows the
behavior of the radial velocity measured for this line at the
different phases of the eclipse. We have obtained the same result
using the Paschen 12 line.  Due to the high excitation potential
of the upper levels NI 8711 \AA\ and Paschen 12 lines there are no
additional absorption components arising in the disk of the secondary.
This is the reason why the behavior of NI and Paschen lines radial velocities is
noticeably different from RV(KI) (see Fig. \ref{3}).

\subsection{The Balmer lines $H_{\alpha} and H_{\delta}$}

Out of eclipse, the $H_{\alpha}$ line demonstrates the absorption profile with two
weak emission peaks at the both sides. Guangwei et al. [18]
attributed formation of these peaks to the rotating
ring-like structure around the primary which can be a
result of the stellar wind. On the base of the radial
velocities changes in the $H_{\alpha}$ line and the orbital velocity of
the F star Chadima et al. [11] argued that the $H_{\alpha}$
emission is formed in the nearest vicinity of the primary.

\begin{figure}[H] \centering
\includegraphics[width=8cm,angle=0]{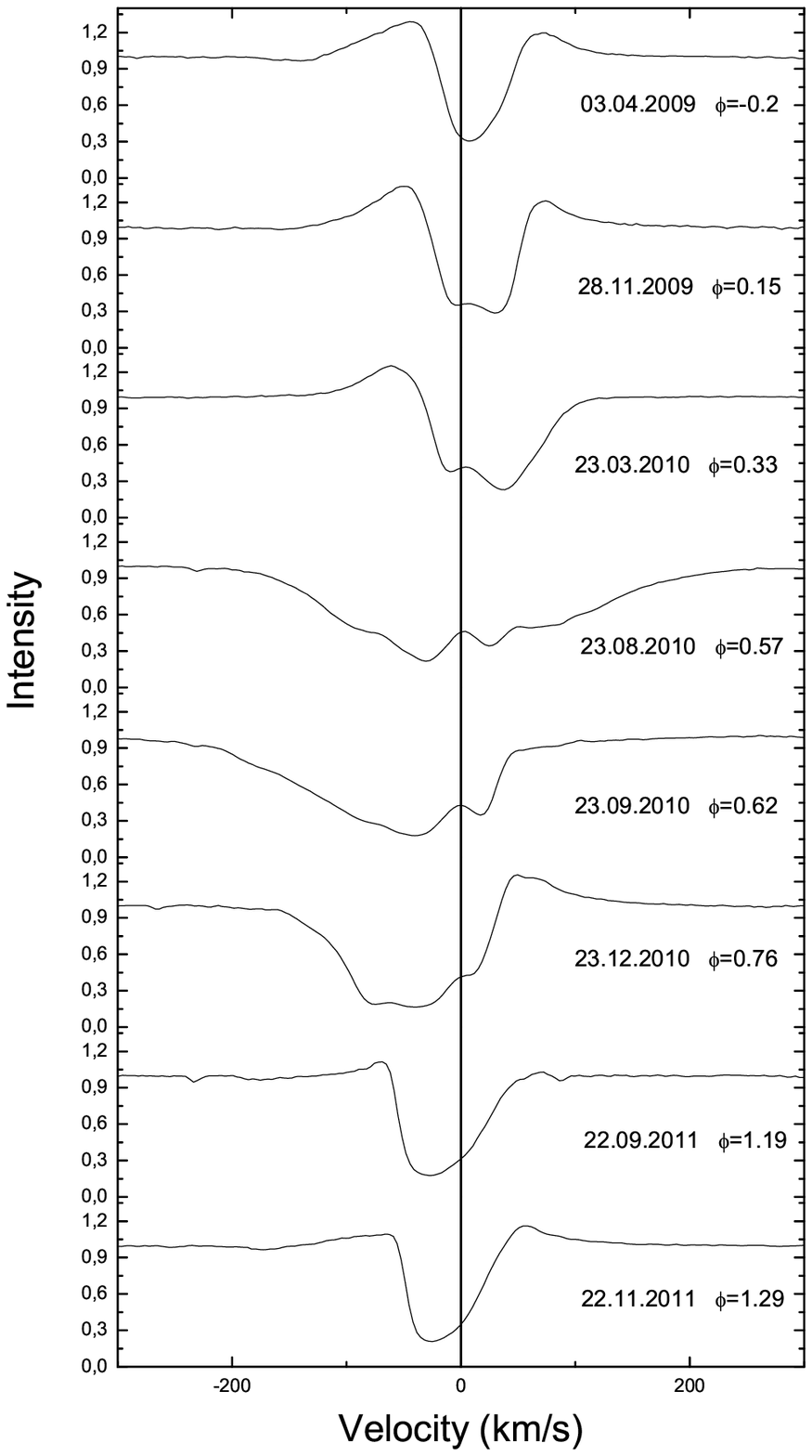}
\caption{The $H_{\alpha}$ line profile variability during the
eclipse.} \label{9}
\end{figure}

Our observations show that the $H_{\alpha}$ line demonstrates
significant changes during the eclipse (Fig. \ref{9}). On the earliest phases of the
eclipse the additional red absorption appeared while the red
emission peak gradually disappeared. At the mid-eclipse phase the
profile is almost symmetric and strongly broad without any signs
of the emission. On the egress phase the picture is unfolded in the
reverse order. The additional absorption shifted to the blue side,
and the blue emission peak disappeared while the red emission
reappeared. At the end of the eclipse the profile returned
to the shape observed out of eclipse.

\begin{figure}[H] \centering
\includegraphics[width=8cm,angle=0]{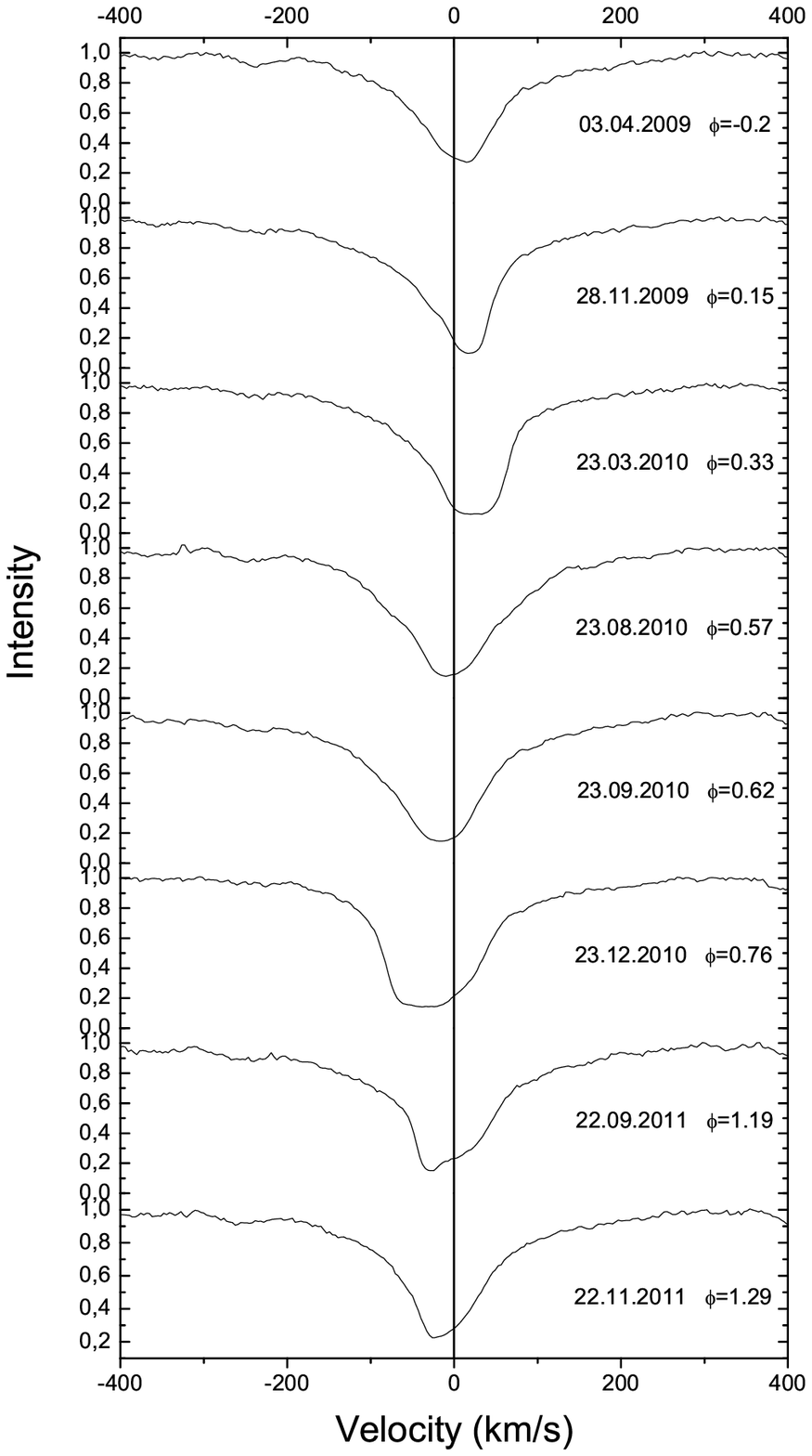}
\caption{The $H_{\delta}$ line profile variability during the
eclipse.} \label{10}
\end{figure}

The $H_{\beta}$ line is strongly blended by the CrII line, so we excluded it
from our analysis of the spectral variability. The $H_{\delta}$
line changes during the eclipse were in general similar to the
$H_{\alpha}$ one. We see, however, no emission components in the
$H_{\delta}$ line, and the line broadening at
the middle of the eclipse was not so strong as in the $H_{\alpha}$ line
(Fig. \ref{10}).


\section{Discussion}

As mentioned in the Introduction, the crucial
question for understanding of $\varepsilon$ Aur evolutionary
status is the problem of the distance to the system. The distance
directly measured by Hipparcos had the error comparable with the
measured value of $D$. The revision of the Hipparcos catalogue
fulfilled by van Leuwen [19] gives the parallax value $\pi= 1.53
\pm 1.29$ that corresponds to the $\varepsilon$ Aur distance $D =
653\pm 551$ pc. Recently other two attempts to determine the
distance to $\varepsilon$ Aur have been undertaken. The first
of them [20] was based on the original method included the
distance calibration as a function of the reddening and 6613 \AA\
DIB intensity. This method was applied to the nearest to the
$\varepsilon$ Aur stars under the assumption about the sufficient
homogeneity of the interstellar medium in the volume investigated.
The authors of the paper [20] give the value $D = 1.5 \pm 0.5$
kpc. However, they have noted, that not very accurate photometric calibration
due to the star pulsations and the influence of the
circumstellar matter in the nearest vicinity of the system could exist.
It should be also noted that taking into account this distance and
the recent interferometric measurements of the diameter of the primary
[1], we obtain the stellar radius in the range of $\sim 200-500
R_\odot$. Such radii are typical for the coldest Mira-like
variables rather than for the F supergiants which radii are
usually $R\sim 80-100 R_\odot$ [21]. The alternative estimate of
$D$ was fulfilled in the PhD thesis by Kloppenborg [1]. This
estimate was based on the astrometric analysis of the plates
from Sproul observatory collection during the 45-year period.
Together with the new orbital parameters of the system, these data
give the distance value $D = 737 \pm 67$ pc. This value is in a
satisfactory agreement with the previous measurement by van de
Kamp [22]. Nevertheless, this estimate was affected by the
uncertainties in the astrometric parameters of the reference stars and
insufficient precision of the orbital parameters. In the
case when only one component of the binary system is observed it
is really a difficult task.

A similarity in the behavior of the radial velocity and the equivalent width
during the current and previous eclipses argue in favor
of the general stability of the disk around the invisible
companion. Using the KI 7699 \AA\ radial velocity measurements and
the velocity of the disk motion ($V_d \approx
30$ km/s) relatively to the F star recently obtained from the interferometric
measurements [1,5] one can reconsider the mass of the
secondary $M_2$. For this purpose, we adopt the distance $D=737$ pc
[1] as the more probable value. Combining the velocity 30 km/s with the
duration of the photometric eclipse from Table 1 we obtain the
estimate of the disk diameter $D_{disk}\sim 12.5$ AU. The
semi-amplitude of the KI 7699 \AA\ radial velocity curve based on
averaging our and Lambert \& Sawyer [8] data is $K=29.7$ km/s.
Adopting this value as the Keplerian velocity at the distance from
the unseen component $0.5 D_{disk}$, we obtain $M_2=6.2 M_{\odot}$.
Substituting this value in the mass function equation, we obtain the
mass of the primary: $M_1=3.5 M_{\odot}$. Thus, the observations during
the 2009-2011 eclipse confirmed the mass estimate by Lambert \&
Sawyer [8] and argue in favor of the 'low-mass' scenario.

It should be noted that observed rotation velocity is probably not
the strictly Keplerian one. Besides, when we measure the line
shift due to the disk rotation we observe the matter motion effect
along the line of sight. This means that the resulting value of the
velocity will be slightly less than the 'true' Keplerian velocity
(at the disk radius). Correspondingly, the true mass can be
slightly larger.

The low-mass scenario is also supported  by Hoard et al. (2010). These authors'
have shown that observed spectral energy distribution of $\varepsilon$ Aur can be
reproduced by the model with masses 2.2$M_\odot$ and 5.9$M_\odot$
for the primary and secondary components correspondingly. The
$\varepsilon$ Aur double-mode pulsation period $\approx 67^d$ and
$123^d$ [24] is shorter, than in the case of the well-studied
normal supergiant $\rho$ Cas ($\approx$ 1 year)[8]. At the same
time the analysis of the chemical composition of $\varepsilon$ Aur
[25] demonstrates its similarity to the massive supergiants
composition, but not to the post-AGB stars. It still seems
difficult to make the final choice between two models without the more
precise distance to the system.

For complete understanding of the $\varepsilon$ Aur system it is very
important to explain the possibility of the origin of the $H_{\alpha}$
line profile broadening up to 200 km/s  near to the
mid-eclipse. As was proposed in one of the earliest models,
the disk around the secondary had the central hole [26]. It was
necessary for the explanation of the little brightening in the
light curve near the middle of the eclipse. However, in the disk
model by Kloppenborg et al. [1,5], this hole is absent. Possibly,
the sensitivity of their method was not enough to resolve this
structure in the disk. However, it is quiet possible that this
central hole is absent or can not be observed due to the edge-on
orientation of the disk.

In this case, a projection of the dusty disk on the sky plane differs
from the ellipse. According to the Lissauer et al. model [27], it
could be something like the flared disk. In this model, the area of
the $H\alpha$ line broadening (Section 3.6.) locates close to the
central region of the disk. It could be structure like the rapidly
rotating hot halo above the disk plane or the magneto-centrifugal
disk wind (if the disk has the magnetic field). In the Section 3.3
we also mentioned the lag between the changes in the FeII and the KI
line profiles. This observational fact means the presence of the
radial temperature gradient in the halo, mentioned above. In the
next paper we are going to consider this effect in more detail.

\section{Conclusion}

The results of the spectroscopic observations during the last
eclipse of $\varepsilon$ Aur demonstrate a stability of the
gaseous component of the disk around the secondary from eclipse to
eclipse. The KI 7699 \AA\ radial velocity curve is in a good
agreement with the previous observations by Lambert \& Sawyer [8],
and argue in favor of the 'low-mass' model of $\varepsilon$ Aur.
The $H\alpha$ line profile broadening is the evidence of the presence
of the hot rapidly rotating gas above the cold dusty disk. Its
temperature increases toward the center. Projection of this area
covers a significant part of the F star disk. The excitation
state of the atomic levels is such that the gas can absorb the
radiation of the primary at the $H\alpha$ line frequencies but it does
not contribute to the emission in this line. This means that the
excitation temperature in the $H\alpha$ line is lower than the
$\varepsilon$ Aur effective temperature (approximately 8000 K).
Theoretical modelling of this region can reveal the source of its
heating.

We thank the Terskol Observatory staff for the observations. This
work was supported in part by the Basic Research Program P21 of
the Presidium of the Russian Academy of Sciences, grant N.Sh.
1625.2012.2, and grant № 1.2.1 of Federal Targeted program
'Science and Science Education for Innovation in Russia 2012-2013'.\\

\clearpage \vspace{1cm}
\begin{center}
\LARGE {\bf Literature}
\end{center}
%
1.  B. Kloppenborg, Ph.D. thesis, University of Denver (2012). \\
2. R.P. Stefanik, G. Torres, J. Lovegrove, V.E. Pera, D.W.
    Latham, J. Zajac,  T. Mazeh, Astrophys. J., 139, 1254  (2010)\\
3.  E.F. Guinan, L.E. DeWarf, in ASP Conf. Series, Vol. 279,
Exotic Stars
as Challenges to Evolution, ed. C. A. Tout \& W. van Hamme (San Francisco: ASP), 121 (2002) \\
4.  S.-S. Huang,  Astrophys. J., 141, 976 (1965) \\
5.  B. Kloppenborg, R.E. Stencel, J. Monnier et al., Nature, \textbf{464}, 870 (2010)\\
6.  G. P. Kuiper, O. Struve, B. Stromgren, Astrophys. J., \textbf{86}, 570 (1937) \\
7. O. Struve, H. Pillans, V. Zebergs, Astrophys. J., \textbf{128}, 287 (1958) \\
8.  D.L. Lambert, S.R. Sawyer, PASP, \textbf{98}, 389 (1986)\\
9. S. Ferluga, D. Mangiacapra, Astron. and
Astrophys, \textbf{243}, 230 (1991) \\
10. M. Saito, S. Kawabata, K. Saijo, H. Sato, Publ. Astron. Soc. Japan, \textbf{39}, 135 (1987)  \\
11. P. Chadima, P. Harmanec, P.D. Bennett et al., Astron. and
Astrophys, \textbf{530}, A146 (2011b)\\
12. P. Chadima, P. Harmanec, S. Yang et.al., IBVS, 5937, 1 (2010)\\
13. R.E. Stencel, JAAVSO, vol. 40 (2012)\\
14. D. Tody, "IRAF in the Nineties" in Astronomical Data Analysis
    Software and Systems II, A.S.P. Conference Ser., Vol 52,
    eds. R.J. Hanisch, R.J.V. Brissenden, J. Barnes, 173 (1993)\\
15. G.A. Galazutdinov, SAO Preprint, 92 (1992)\\
16. Welty, D.E., Hobbs, L.M., ApJS, \textbf{133}, 345 (2001)\\
17 N. Mastrodemos, M. Morris, Astrophys. J. \textbf{497}, 303 (1998) \\
18. C. Guangwei, T. Huisong, X. Jun, L. Yongsheng, Astron. and
Astrophys, \textbf{284}, 874 (1994)\\
19.  F. van Leeuwen, Astron. and Astrophys, \textbf{474}, 653 (2007b)\\
20. E.F. Guinan, P. Mayer, P. Harmanec, et al., Astron. and
Astrophys, \textbf{546}, A123 (2012)\\
21. Th. Schmidt-Kaler , The Physical Parameters of the Star, in:
Landolt-Bornstein, ed. K.-H. Hellwege, New Series Vol. VI, 2b,
Springer, Berlin, Heidelberg, New York, p.1ff (1982)\\
22. P. van de Kamp, Astron. J. , \textbf{83} , 975 (1978)\\
23.  D. W. Hoard, S. B. Howell, R. E.  Stencel, Astrophys. J., \textbf{714}, 549 (2010) \\
24. H. Kim, JASS, \textbf{25}, 1 (2008)\\
25.  K. Sadakane, E. Kambe, B. Sato, S. Honda, O. Hashimoto, Publ. Astron. Soc. Japan, \textbf{62}, 1381 (2010)\\
26. Wilson, R.E., Astrophys. J. \textbf{170}, 529 (1971)\\
27. J.J.\,Lissauer, S.J.\,Wolk, C.A.\,Griffith, D.E.\,Backman ,
Astrophys. J., \textbf{465}, 371 (1996) \\

\end{document}